\documentclass[english,aip,jcp,numerical,preprint]{revtex4-1}
\usepackage{hyperref}
\usepackage[T1]{fontenc}
\usepackage[latin9]{inputenc}
\usepackage{float}
\usepackage{amsmath}
\usepackage{amssymb}
\usepackage{wasysym}
\usepackage{graphicx}
\usepackage{esint}
\usepackage{natbib}
\usepackage{graphicx}


\usepackage{babel}

\begin{document}

\preprint{This line only printed with preprint option}

\title{Critical exponents of colloid particles in bulk and confinement}

\author{Helge Neitsch}
\email{helge.neitsch@physik.tu-berlin.de}
\author{Sabine H. L. Klapp}
\email{klapp@physik.tu-berlin.de}
\affiliation{
	Institut für Theoretische Physik, Sekr. EW 7-1,
	Fakult\"at II für Mathematik und Naturwissenschaften,
	TU Berlin, Hardenbergstraße 36, D-10623 Berlin
}

\begin{abstract}
	Using grand canonical Monte Carlo simulations, we investigate the
	percolation behavior of a square well fluid with an ultra-short range
	of attraction in three dimension (3D) and in confined geometry. The
	latter is defined through two parallel and structureless walls (slit-pore).
	We focus on temperatures above the critical temperature of the (metastable)
	condensation transition of the 3D system. Investigating a broad range
	of systems sizes, we first determine the percolation thresholds, i.~e.,
	the critical packing fraction for percolation $\eta_{c}$.
	For the slit-pore systems, $\eta_{c}$ is found to vary with the wall
	separation $L_{z}$ in a continuous but non-monotonic way,
	$\eta_{c}(L_{z}\rightarrow\infty)=\eta_{c}^{\text{3D}}$.
	We also report results for critical exponents of the percolation transition,
	specifically, the exponent $\nu$ of the correlation length $\xi$
	and the two fisher exponents $\tau$ and $\sigma$ of the cluster-size
	distribution. These exponents are obtained from a finite-size analysis
	involving the cluster-size distribution and the radii of gyration
	distribution at the percolation threshold. Within the accuracy of
	our simulations, the values of the critical exponents of our 3D system
	are comparable to those of 3D random percolation theory. For narrow
	slit-pores, the estimated exponents are found to be close
	to those obtained from the random percolation theory in two dimensions.
\end{abstract}
\maketitle

\section{Introduction}

Percolation is a geometrical transition, in which interacting units such as
the particles in a fluid, the spins on a lattice, or the nodes of a network
spontaneously form system-spanning clusters termed the ``percolated phase''.\cite{IntPerThe}
In contrast the units are distributed homogeneously or form isolated clusters of finite size
in the non-percolated phase.

In this publication we focus on continuous systems, where percolation was originally discussed
as a phenomenon in the context of flow through porous media. Later, the
statistical theory of percolation has been used to understand the critical
behavior of fluid (Fisher's droplet model) \cite{Fisher1970a,Fisher1970}, particularly the divergence
of the correlation length and the critical exponents characterizing the 
behavior close to the critical point. Indeed, it is well known that
percolation obeys concepts of universality and scaling, similar to second-order
thermodynamic phase transitions.\cite{Sator2003} This includes
the universality class of the so-called random percolation, which is assumed
to apply to continuous systems such as fluids.

More recent (experimental and theoretical) research on continuous percolation
often involves colloidal suspensions. One important topic in this area concerns
the percolation of rod-like colloids, prominent examples being carbon
nanotubes and other carbon-based particles \cite{Otten2012,Otten2011,Otten2009,Schilling2007}.
From an applicational point of view the underlying idea is that the percolated network leads to
lightweight materials with strongly enhanced mechanical stability and electrical
(and/or thermal) conductivity. Another main topic, which concerns particularly
complex colloidal mixtures \cite{Zac2008,Duda2009} and colloids with directional interactions \cite{Rovigatti2012,Romano2011,Sciortino2011},
is the intimate relation between percolation and the formation of a physical gel,
which is a state in which particles are connected via bonds of limited lifetime.
It is now well established that such colloidal gels, which are characterized
through a very specific dynamic behavior, can form at extremely low
packing fractions. Note, however, that gelation is a phenomenon which normally
occurs for very strong coupling conditions, i.~e., at temperatures far
below those related to the vapor-liquid critical point (if the latter exists at all).

In the present study we are interested in the characteristics
of the percolation transition at moderate (supercritical) temperatures,
focusing on a system of spherical colloids with attractive interactions.
Specifically, we aim to determine the percolation threshold, the scaling behavior
and the related critical exponents. We consider both, three-dimensional (3D)
systems and systems in slit-pore geometries which are infinite in only two
dimensions (2D). Our investigations are based on grand-canonical Monte Carlo
(GCMC) computer simulations combined with a finite-size scaling analysis.

Our study is partially motivated by a series of recent Monte Carlo (MC) results by Nezbeda
and coworkers.\cite{Skvor2007,Skvor2009,Skvor2011}These authors investigated
the percolation in the corresponding supercritical regimes of various 3D (bulk) continuous model systems, differing in the precise form
of the interaction potential.\cite{Skvor2009}
Using different geometric cluster definitions and spanning rules,
they found perfect scaling behavior for all systems investigated. However,
the critical exponents turned out to be strongly dependent on the specific
interaction potential and the temperature. This obviously contradicts the
expectation that the percolation transition in these systems is universal.
Indeed, in ideal random percolation the critical exponents and scaling functions
only depend on the dimensionality and the symmetries of the system.\cite{IntPerThe}

Here we present results from a GCMC study involving colloids with an ultra-short
range of attraction, modeled by a square-well potential with an
attraction width of only four percent of the colloidal diameter. This is
considerably smaller than the range of fifty percent studied in Ref.~\onlinecite{Skvor2009}.
Interestingly, our GCMC results for the 3D case do indicate \textit{universality}
of the exponents in the sense that they do not depend on the temperature in the range considered.
The exponents do, however, depend on the
spatial dimensionality, consistent with the predictions from random percolation theory.

We note that our model is by no means artificial. Indeed, colloids with
ultra-short ranged interactions can be experimentally realized
by adding small, non-adsorbing polymers to the colloidal solution, yielding
short-range, attractive depletion interactions of tunable range \cite{PeterJ.Lu2008}.
Another, recently studied example (where the justification of the ultra-short
ranged square-well model has actually been checked using scattering data)
are silica nanoparticles with added lysozyme \cite{Bharti2011}.

This paper is organized as follows. In
Sec.~\ref{sec:Model} we introduce the model and discuss its bulk phase diagram.
Numerical details of the GCMC simulations are described in Sec.~\ref{sec:Computational}.
In Sec.~\ref{sec:Results} we present our numerical results for the percolation
thresholds and critical exponents of the bulk and slit-pore systems.
Specifically, we consider the exponents of the correlation length,
the cluster size distribution, and the radius of gyration.
In Sec.~\ref{sub:Confinement-induced-shift} we finally present results for the 
confinement-induced shift of the percolation threshold, and we provide
a simple theory based on scaling arguments to describe this shift.
We close this paper with the conclusions in Sec.~\ref{sec:Conclusion}.

\section{Model\label{sec:Model}}

In our colloidal model system, the particles interact via an attractive
square-well (SW) potential defined as 
\begin{equation}
	u_{\text{SW}}(\boldsymbol{r}_{ij})=\begin{cases}
	\infty, & r_{ij}<\sigma\\
	-\epsilon, & \sigma\le r_{ij}<\lambda\\
	0, & \lambda\le r_{ij}
	\end{cases}\text{.}\label{eq:def_sw}
\end{equation}
In Eq.~\eqref{eq:def_sw} $r_{ij}=|\boldsymbol{r}_{ij}|=|\boldsymbol{r}_{i}-\boldsymbol{r}_{j}|$
is the particle distance, $\sigma$ is the hard-core diameter, $\epsilon>0$
is the strength of the attraction and $\lambda$ determines the range
of this attraction, i.~e., $\lambda>\sigma$. Here we choose a value
$\lambda=1.04\sigma$, corresponding to a system with an ultra-short
range of attraction.

The SW model with ultra-short ranged attraction has attracted growing
attention, since it is known to be a simple but adequate model for
colloidal suspensions. Indeed, short-ranged attractive colloids can
be realized experimentally by adding small non-adsorbing polymers
to the colloidal solution. This creates a depletion effect leading
to an effective attraction between the colloids. To screen the remaining
(attractive) Van der Waals interactions between the colloids, one
chooses a solvent of similar dielectric permittivity.\cite{Hoog2001,Wijting2003,Zac2008}
If the difference in the length scales between the polymers and the
colloids is sufficiently large, the degrees of freedom associated
with the polymers can be ``integrated out'' of the theoretical description.
The remaining implicit effect is an effective short-ranged attraction
between the colloids due to depletion interactions.\cite{Asakura1958}
Regarding the actual value of $\lambda$, we note that for values $\text{\ensuremath{\lambda}}\lesssim1.25\sigma$
the stable liquid phase disappears and a typical colloidal phase diagram
involving only a fluid and a solid phase arises.\cite{NeerAsherie1996}

We consider our colloidal model in three dimensions (bulk) and in
slit-pore geometries. The latter are realized by two plane-parallel
structureless walls, which are parallel to the x-y-plane of the coordinate
system and located at $\pm L_{z}/2$. The wall-particle interaction
is modeled via
\begin{equation}
	u_{\text{wall}}(d)=\begin{cases}
	\infty, & d<\sigma/2\\
	0, & d\geq\sigma/2
	\end{cases}
	\label{eq:def_wall}
\end{equation}
with the particle-wall distance $d$. Thus, the slit-pore-particle
interaction is given by
\begin{equation}
	u_{\text{slit}}(\boldsymbol{r}_{i})=u_{\text{wall}}(L_{z}/2+z_{i})+u_{\text{wall}}(L_{z}/2-z_{i}))
	\label{eq:def_slit}
\end{equation}
where $z_{i}$ is the z-component of the center of mass position
$\boldsymbol{r}_{i}$ of the $i$th particle. Such a purely repulsive
colloid-wall potential can be realized experimentally by using surface
structures like a polymer-coated brush, which is penetrable for the
polymers but not for the colloids.\cite{Wijting2003,Schmidt2004}

Recent research has shown that short-ranged systems, such as the present
SW model, exhibit thermodynamic properties which are insensitive to
the specific shape of the interaction potential and approximately
fulfill an extended law of corresponding states.\cite{Noro2000,Foffi2006,J.Largo2008}
This will be useful to derive a rough draft of the bulk phase diagram
of our system. To this end we first consider a limit case of the SW
model, the Baxter model, which is also referred to as the
adhesive spheres (AHS) model. Within this model the limits $\lambda\rightarrow\sigma$
and $-\epsilon\rightarrow-\infty$ are determined while keeping the ratio of the
second virial coefficient of the AHS model, $B_{2}^{\text{AHS}}$,
to the second virial coefficient of the hard sphere system
(HS), $B_{2}^{\text{HS}}=4\pi\sigma^{3}/3$, constant, that
is 
\begin{equation}
	\frac{B_{2}^{\text{AHS}}}{B_{2}^{\text{HS}}}=4\tau^{\text{AHS}}-1\overset{!}{=}const\text{.}
	\label{eq:tau_ahs}
\end{equation}
Equation~\eqref{eq:tau_ahs} defines the stickiness parameter $\tau^{\text{AHS}}$,
which describes to which extend the AHS particles tend to glue together.
Hence $\tau^{\text{AHS}}$ acts as an effective dimensionless temperature
in the Baxter model. 

The Baxter model has been extensively investigated both by theory and by simulation.\cite{Baxter1968a,Miller2004,J.Largo2008}
Within the fluid phase regime it exhibits two types of phase transitions,
namely a percolation transition and a vapor-liquid (vl) condensation
transition.
The critical parameters of the latter have been determined with very high accuracy
using GCMC simulations.
Their values are
$\rho_{c,\text{vl}}^{\text{AHS}}=0.508(10)/\sigma^{3}$
(the critical particle density) and $\tau_{c,\text{vl}}^{\text{AHS}}=0.1133(5)$
(the critical temperature). Based on these values, we can now estimate
the locus of the vapor-liquid critical point in the present square-well
system. The first assumption is that the packing fraction $\eta=\pi\sigma^{3}\rho/6$
associated to the critical density remains constant, when we replace
$\sigma$, which corresponds to the average distance of two bonded
particles in AHS system, with the corresponding average distance $d_{\text{av}}$
in the SW system. In other words, we require \cite{J.Largo2008,Foffi2007}
\begin{equation}
	\rho_{c,\text{vl}}d_{\text{av}}^{3}\overset{!}{=}\rho_{c,\text{vl}}^{\text{AHS}}\sigma^{3}\text{.}
	\label{eq:deduce_map_dens}
\end{equation}

For the present model, see Eq.~\eqref{eq:def_sw}, we have $d_{\text{av}}=\sigma+(\lambda-\sigma)/2\text{.}$
Inserting this into Eq.~\eqref{eq:deduce_map_dens}, one obtains 
\begin{equation}
	\rho_{c,\text{vl}}=\frac{\sigma^{3}\rho_{c,\text{vl}}^{\text{AHS}}}{(\sigma+(\lambda-\sigma)/2)^{3}}\text{.}
	\label{eq:dens_ahs}
\end{equation}
To estimate the critical temperature, we apply the Noro-Frenkel law
of corresponding states.
It states that two systems which are similar in nature by having both a hard-core repulsion and an (ultra-)short-ranged attraction
have equal reduced second virial coefficients, provided we reduce with $B_{2}^{\text{HS}}$ close to the critical temperatures in each case.\cite{Noro2000}
This allows us to apply a mapping of the phase diagram of one of those systems to the other.
Taking the AHS as the reference system and equating its reduced
second virial coefficient with the reduced second virial coefficient
of a short-ranged SW model yields 
\begin{equation}
	\frac{B_{2}^{\text{AHS}}}{B_{2}^{\text{HS}}}=4\tau^{\text{AHS}}-1\overset{!}{=}\frac{B_{2}^{\text{SW}}}{B_{2}^{\text{HS}}}=1-(\left(\lambda/\sigma\right)^{3}-1)(e^{1/T^{*}}-1)\text{.}
	\label{eq:equat_scn_virial}
\end{equation}
In Eq.~\eqref{eq:equat_scn_virial} we introduced the reduced temperature
$T^{*}=k_{B}T/\epsilon$. Solving Eq.~\eqref{eq:equat_scn_virial}
with respect to $T^{*}$ we obtain
\begin{equation}
	T^{*}=\left[\ln\left(1+\frac{1}{4\tau^{\text{AHS}}\left((2-\lambda/\sigma)^{-3}-1\right)}\right)\right]^{-1}.
	\label{eq:temp_ahs}
\end{equation}
Based on Eqs.~\eqref{eq:dens_ahs} and \eqref{eq:temp_ahs}, we can
estimate the critical parameters of the present model by inserting
the known values for $\rho_{c,\text{vl}}^{\text{AHS}}$ and $\tau_{c,\text{vl}}^{\text{AHS}}$,
respectively. This yields 
\[
	\eta_{c,\text{vl}}=0.250(5)\text{, }T_{c,\text{vl}}^{*}=0.346(5)
\]
where $\eta_{c,\text{vl}}$ is the critical packing fraction of the
vapor-liquid condensation transition.

In addition to determining the critical point, we may use
the mapping procedure described above to obtain an estimate of the
full fluid-fluid coexistence curve (binodal), as reviewed in Ref.~\onlinecite{Zaccarelli2007}.
To this end, we apply Eq.~\eqref{eq:dens_ahs} and Eq.~\eqref{eq:temp_ahs}
to GCMC data for the binodal taken from Fig.~3 of Ref.~\onlinecite{Miller2003}.
Moreover, from integral equation studies \cite{Foffi2002} of a SW
system with $\lambda=1.03\sigma$, the fluid-solid coexistence line is
known (see Fig.~9 of Ref.~\onlinecite{Foffi2002}). Taking their data and
applying the mapping procedure again, we obtain an estimate of the
fluid branch of the fluid-solid coexistence line in our system. Finally,
we also mapped the data of the percolation threshold given in Ref.~\onlinecite{Miller2003}.

\begin{figure}[h]
	\includegraphics[]{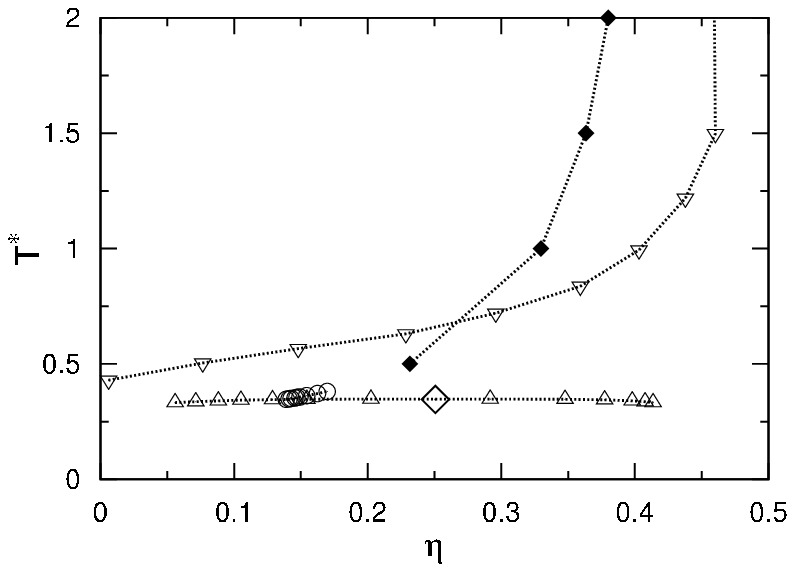}
	\caption{Estimated phase diagram of the bulk system, where we applied the mapping
		formulas (\ref{eq:dens_ahs}) and (\ref{eq:temp_ahs}) (empty symbols).
		$\Diamond\text{:}$ the estimated critical point at $\eta_{c,\text{vl}}=0.250(5)$
		and $T_{c,\text{vl}}^{*}=0.346(5)$, $\vartriangle\text{:}$ the mapped
		metastable fluid-fluid coexistence line, $\circ\text{:}$ the mapped
		percolation line and $\triangledown\text{:}$ the mapped fluid branch
		of the fluid-crystal coexistence line. Our present results for the
		percolation threshold are indicated by filled symbols. All dotted
		lines are guides to the eye.
		\label{fig:Fig1}}
\end{figure}

Since we are partly far away from the critical point during such
mapping, some doubts are appropriate regarding the justification of
that strategy. Nevertheless, the mapping results give a first idea
of the phase diagram of the present system. A summary of the results
of these mapping procedures is given in Fig.~\ref{fig:Fig1}.
Included in this figure are our present GCMC results for the bulk
percolation thresholds, which will be discussed in more detail in
Sec.~\ref{sec:Results}. The resulting percolation line divides the
phase diagram into two regions characterized by finite (low~$\eta$)
and infinite (high~$\eta$) clusters, respectively. Note that in
both cases the clusters are \textit{transient} in character, since
there are no permanent bonds in our system.

\begin{figure}[h]
	\includegraphics[]{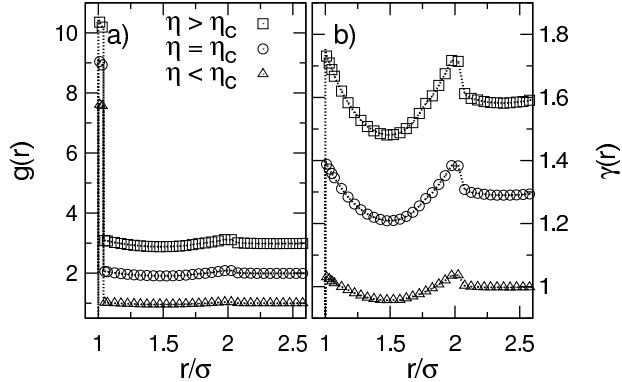}
	\caption{a)~Radial distribution function $g(r)$ and b)~cavity function $\gamma(r)$
		at $T^{*}=0.5$ for packing fractions below, at and above the percolation
		threshold {[}$\eta\approx0.1\mbox{ (\ensuremath{\Square}), }\eta=\eta_{c}\approx0.23\text{ (\ensuremath{\Circle}) and }\eta\approx0.3\text{ (\ensuremath{\triangle})}${]}.
		To improve the visibility, the graphs for $\eta=\eta_{c}$ ($\eta>\eta_{c}$)
		are shifted upwards by $1.0$ ($2.0$) in a) and by $0.3$ ($0.6$)
		in b). Dotted lines are guide to the eye.
		\label{fig:Fig2}}
\end{figure}

According to Fig.~\ref{fig:Fig1}, the lowest temperature
$T^{*}=0.5$, for which we investigated the percolation transition,
lies near or even within the metastable region between fluid and solid
states. Therefore we carefully examined the packing fraction and configurations
generated during each GCMC run in order to guarantee that we average only in one phase.
In particular, we analyzed the structure
in terms of the radial distribution function $g(r)$. In Fig.~\ref{fig:Fig2}
we show $g(r)$ together with the cavity function $\gamma(r)=g(r)e^{-u_{\text{SW}}(r)/k_{B}T}$
of the bulk system at $T^{*}=0.5$ at three packing fractions below,
at and above the critical packing fraction for percolation $\eta_{c}$,
which will be defined in Sec.~\ref{sec:Results}. Since $g(r)$
and $\gamma(r)$ show no indication of any long rang order, we may
conclude that we are indeed in the fluid phase.

\section{Computational details\label{sec:Computational}}

To investigate the percolation behavior, we carried out standard GCMC
simulations.\cite{ComSimLiq} For the bulk system calculations have
been performed at several reduced temperatures $T^{*}=0.5,\,1.0,\,1.5,\,2.0$.
For the confined system we considered one fixed temperature, $T^{*}=0.5$,
and several wall distances ranging from $L_{z}=1.5\sigma$ to $L_{z}=10.0\sigma$.
The systems are equilibrated by performing at least $10^{5}$ GCMC
cycles, where each cycle consist on $\left\langle N\right\rangle $
single particle displacement, insertion or deletion attempts, with
$\left\langle N\right\rangle $ being the average number of particles
in the system. Ensemble averages are obtained by analyzing at least
$10^{5}$ equilibrium configurations, separated by $80$ cycles. With
this choice of parameters the energy autocorrelation function between two averaging
events is close to zero in all cases.

To determine the percolation threshold we performed
at least four simulation runs at different basis lengths $L$ for each system, where
we have $L_{x}=L_{y}=L_{z}=L$ for the bulk system and $L_{x}=L_{y}=L>L_{z}$
for the slit-pore system. In each series, we chose the smallest $L$
such that $\left\langle N\right\rangle \approx500$ at $\eta=0.1$.
The largest systems we investigated contained about to $30000$ particles.

An important aspect of any simulation study targeting percolation
properties concerns the definition of clusters. Here we use a simple
configurational criterion. Since we are dealing with a square-well
model, two particles at positions $\boldsymbol{r}_{i}$ and $\boldsymbol{r}_{j}$
are unambiguously defined as connected if their distance is within
the attractive well of the pair potential, i.~e., $r_{ij}<\lambda\text{.}$
The resulting configurational clusters may be considered as ``percolated''
if they are connected to their own periodic image.\cite{Edvinsson1999,Skvor2009}
In the 3D (bulk) system, we allow for percolation
in any of the three spatial directions. The slit-pore system, on the
other hand, is periodic only in the x- and the y-direction. Hence
percolation can only occur in these two dimensions (2D). The whole
system is considered as percolated if at least one percolating cluster
exists. To identify clusters and check for percolated configurations
in our continuous model system, we implemented an algorithm as described
in Ref.~\onlinecite{Skvor2009}. 

\section{Analysis of the percolation transition\label{sec:Results}}

\subsection{Determination of the percolation threshold\label{sub:percolation_threshold}}

The purpose of this section is to outline our approach to determine
the percolation threshold. Indeed, the definition of percolation itself
is not unique in the literature. It depends on the system to
be investigated, the definition of a percolated state and the definition
of bonds.\cite{IntPerThe,Skvor2007,HeyMel1989} This may lead to
incompatible results for a single system. On the other hand, the definition
of the percolation \textit{threshold} is much clearer, at least for
lattice systems. There one typically considers the probability $p$
of finding a given lattice site occupied. The threshold $p_{c}$ is
then defined by the following criterion, that
``for $p$ above $p_{c}$ one percolating network exists; for $p$
below $p_{c}$ no percolating network exists.''\cite{Stauffer1979}

We now have to clarify which quantity in our continuous system should
play the role of the control variable $p$. Since we perform
GCMC simulations, a natural control variable would be the activity
(or the chemical potential, respectively). Here we prefer to
use the (average) packing fraction $\eta$ due to the obvious analogy
between $\eta$ and the occupancy of lattice sites.\cite{Skvor2009}

To determine the percolation threshold $\eta_{c}$, we examine the
probability $\Pi(\eta,L)$ of finding a spanning cluster in a given
system (3D or slit-pore) with a characteristic length $L$ at packing
fraction $\eta$. This probability, which depends on the system size
$L$ and the packing fraction $\eta$, can be calculated via an ensemble
average of a test function $\pi(\{\boldsymbol{r}_{k}\},L)$ 
\begin{equation}
	\Pi(\eta,L)=\left\langle \pi(\{\boldsymbol{r}_{k}\},L)\right\rangle \text{,}
	\label{eq:def_PI}
\end{equation}
where $\pi(\{\boldsymbol{r}_{k}\},L)=1$ if the configuration defined
by $\{\boldsymbol{r}_{k}\}$ has one spanning cluster and $0$ if
all clusters are finite. It is commonly assumed (not rigorously proven)
that $\Pi(\eta,L)$ fulfills a single-variable scaling law \cite{Stauffer1979,IntPerThe}
of the form 
\begin{equation}
	\Pi\left(\eta,L\right)=\tilde{\Pi}\left(x\right)=\tilde{\Pi}\left((\eta-\eta_{c})L^{1/\nu}\right)\text{.}
	\label{eq:Pi}
\end{equation}
The ansatz \eqref{eq:Pi} involves the percolation threshold $\eta_{c}$,
the critical exponent $\nu$ to be defined later {[}see Eq.~\eqref{eq:pow_law_cor_length}{]}
and the master curve $\tilde{\Pi}(x)$, which is assumed to be unique
for each set of parameters. To obtain an accurate estimate of $\eta_{c}$
we employ the following strategy (for a more detailed description
see~Ref.~\onlinecite{IntPerThe}).

\begin{figure}[h]
	\includegraphics[]{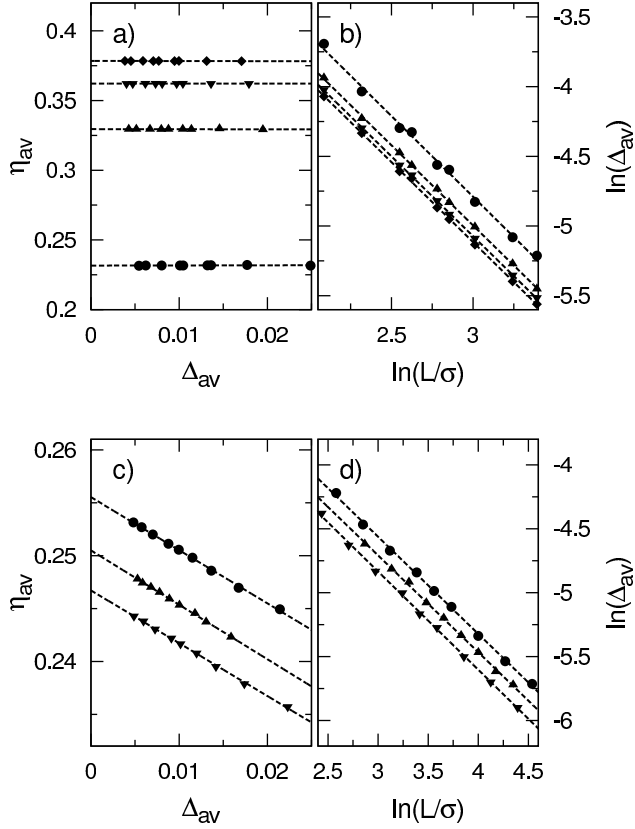}
	\caption{Upper panel: Determination of the percolation threshold and the exponent
		$\nu$ for bulk systems of temperatures
		$T^{*}=0.5\text{ (\ensuremath{\bullet})},\,1.0\text{ (\ensuremath{\blacktriangle})},\,1.5\text{ (\ensuremath{\blacktriangledown}) and}\,2.0\text{ (\ensuremath{\blacklozenge})}$.
		The plots show a) the average packing fraction $\eta_{\text{av}}$
		as function of the standard deviation $\Delta_{\text{av}}$ and b)
		the (logarithm of) the standard deviation as function of system size.
		The dashed lines are obtained from least square fit. From a) the percolation
		threshold follows as the intercept on the y-axis {[}see~Eq.~\eqref{eq:lin_fit}{]}.
		From b), the exponent $\nu$ follows as the slope of the lines {[}see~Eq.~\eqref{eq:pow_law_del}{]}.
		Lower panel: Analog to the upper panel, but for slit-spore
		systems at temperature $T^{*}=0.5$ and $L_{z}=3.0\sigma\text{ (\ensuremath{\bullet})},\,3.5\sigma\text{ (\ensuremath{\blacktriangle}) and }4.0\sigma\text{ (\ensuremath{\blacktriangledown})}$.
		\label{fig:Fig3}}
\end{figure}

We first note that the probability $\Pi(\eta,L)$ is expected to have
a sigmoidal shape with $\Pi(0,L)=0$ and $\Pi(\eta\rightarrow\eta_{\text{cp}},L)=1$,
where $\eta_{\text{cp}}=\sqrt{2}\pi/3\approx0.74$ is the packing
fraction at close-packing (cp). Given this shape, the first derivative
of $\Pi(\eta,L)$ with respect to $\eta$ is a function with a peak
at the packing fraction characterizing the steepest ascent of $\Pi(\eta,L)$.
This feature, combined with the fact that $\partial\Pi/\partial\eta$
is normalized to one, allows us to define an average density
$\eta_{\text{av}}$ according to 
\begin{equation}
	\eta_{\text{av}}(L)=\int_{0}^{\eta_{\text{cp}}}\! d\eta\,\eta\frac{\partial\Pi}{\partial\eta}\text{.}
	\label{eq:def_eta_av}
\end{equation}
Furthermore, we can define the standard deviation $\Delta_{\text{av}}$ via 
\begin{equation}
	\Delta_{\text{av}}^{2}(L)=\int_{0}^{\eta_{\text{cp}}}\! d\eta\,\left(\eta-\eta_{\text{av}}\right)^{2}\frac{\partial\Pi}{\partial\eta}\text{.}
	\label{eq:def_delta2}
\end{equation}
Using Eq.~\eqref{eq:Pi} to substitute $\tilde{\Pi}$ for $\Pi$ and
writing $\eta=x\, L^{-1/\nu}+\eta_{c}$, we obtain from Eq.~\eqref{eq:def_eta_av}
the linear dependence 
\begin{equation}
	\eta_{\text{av}}=A\, L^{-1/\nu}+\eta_{c}
	\label{eq:pow_law_eta}
\end{equation}
where $A$ is a constant. Applying similar arguments to Eq.~\eqref{eq:def_delta2}
results in 
\begin{equation}
	\Delta_{\text{av}}\propto L^{-1/\nu}\text{.}
	\label{eq:pow_law_del}
\end{equation}
Combining Eqs.~\eqref{eq:pow_law_eta}~and~\eqref{eq:pow_law_del}
finally yields the expression 
\begin{equation}
	\eta_{\text{av}}-\eta_{c}\propto\Delta_{\text{av}}\text{.}
	\label{eq:lin_fit}
\end{equation}
Thus, Eq.~\eqref{eq:lin_fit} predicts a \textit{linear} relationship
between $\eta_{\text{av}}$ and $\Delta_{\text{av}}$, with $\eta_{c}$
playing the role of an y-intercept. We have used this relationship
to obtain the percolation threshold. In order to actually calculate
$\eta_{\text{av}}$ and $\Delta_{\text{av}}$, we have fitted the
numerical data for $\Pi(\eta,L)$ according to the expression (see~Ref.~\onlinecite{Skvor2009})
\begin{equation}
	f(x;\left\{ a_{i}\right\} )=1+\tanh\left(\sum_{i=0}^{5}a_{i}x^{i}\right)\label{eq:def_sigmoid}
\end{equation}
involving the six fitting parameters $\left\{ a_{i}\right\} $. Some
representative numerical results for $\eta_{\text{av}}$ as function
of $\Delta_{\text{av}}$ are given in Fig.~\ref{fig:Fig3}a)~and~c).
In~Fig.~\ref{fig:Fig3}a) we consider bulk
systems at various reduced temperatures. Clearly, the data points
for $\eta_{\text{av}}$ form nearly perfect straight lines,
consistent with Eq.~\eqref{eq:lin_fit}, which allows us to extract
the percolation threshold density $\eta_{c}$ easily. The averaged gradient
$A$ of these nearly parallel lines is essentially zero, $A=-0.002\pm0.005$.
The results for $\eta_{c}$ at the four temperatures investigated
are plotted in Fig.~\ref{fig:Fig1}. Figure~\ref{fig:Fig3}c)
shows the same type of data for the slit-pore systems with various
wall separations. The results reflect a very good accuracy of
the approach again. Here, the averaged gradient is close to $-1/2$
(i.~e., $A=-0.505\pm0.005$). The resulting percolation thresholds
for the confined systems are further discussed in Sec.~\ref{sub:Further-critical-exponents}. 

\begin{figure}[h]
	\includegraphics[]{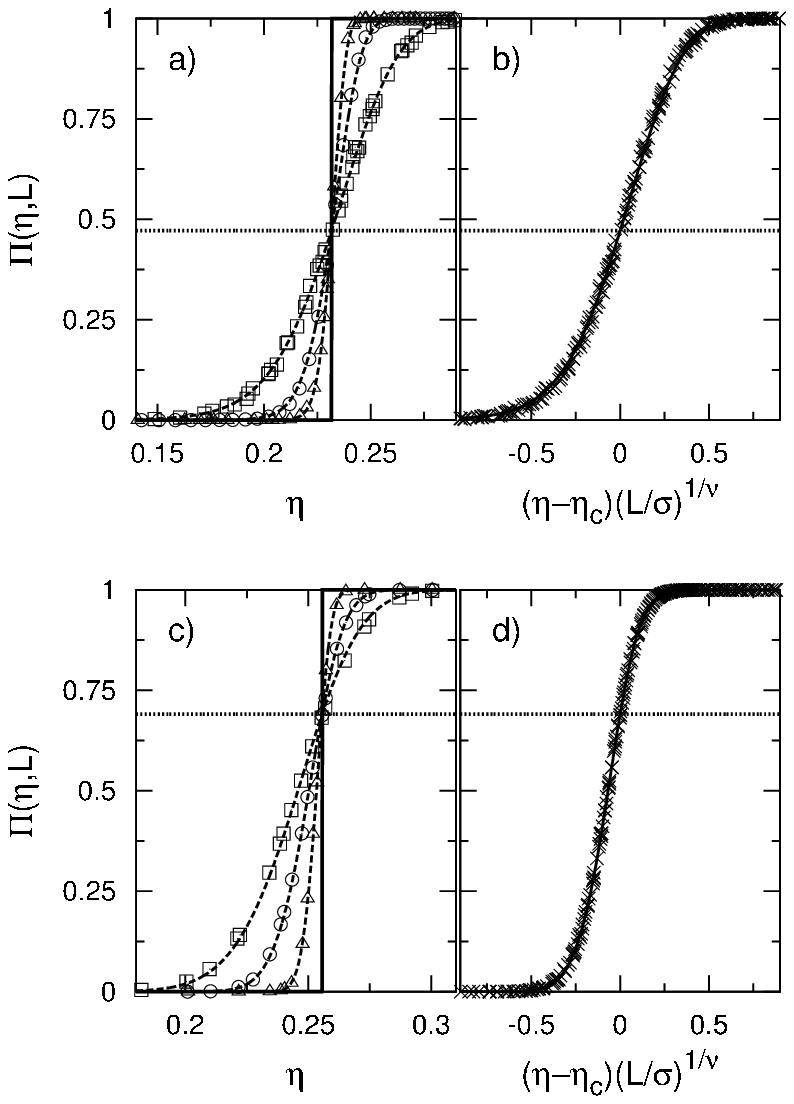}
	\caption{Upper panel: Percolation probability $\Pi(\eta,L)$ of the bulk system
		at $T^{*}=0.5$. a)~Simulation data
		for three system sizes $L=8.06\sigma\text{ (\ensuremath{\square}), }16.12\sigma\text{ (\ensuremath{\bigcirc}) and }29.7\sigma\text{ (\ensuremath{\vartriangle})}$.
		The dashed lines are obtained via least square fits of Eq.~\eqref{eq:def_sigmoid}.
		The bold solid line denotes the assumed limiting shape for $L\rightarrow\infty$.
		The dotted line marks $\Pi(\eta_{c},L)$. b)~Shifted and scaled simulation
		data $\text{(\ensuremath{\times})}$ for all nine investigated system
		sizes, ranging between $L=8.06\sigma$ and $L=29.7\sigma$. The bold
		solid line denote the assumed shape of the master curve $\tilde{\Pi}(x)$
		as obtained from a least square fit of Eq.~\eqref{eq:def_sigmoid}.
		Lower panel: Same as the upper panel, but for a slit-pore system at
		$T^{*}=0.5$, $L_{z}=3.0\sigma$ and three system sizes $L=13.21\sigma\text{ (\ensuremath{\square}), }29.54\sigma\text{ (\ensuremath{\bigcirc}) and }93.42\sigma\text{ (\ensuremath{\vartriangle})}$.
		\label{fig:Fig4}}
\end{figure}

In a similar fashion, we can use Eq.~\eqref{eq:pow_law_del} to obtain
the critical exponent $\nu$ {[}which is involved in the scaling law
\eqref{eq:Pi}{]}. To this end we plot in Fig.~\ref{fig:Fig3}b)~and~d)
the standard deviation $\Delta_{\text{av}}$ as a function of the
system size in a double-logarithmic representation. Again, we find
that the relationship predicted by Eq.~\eqref{eq:pow_law_del} is
essentially perfectly fulfilled for both, bulk and slit-pore systems.
Finally, having determined $\eta_{c}$ and $\nu$, we can test the
underlying scaling assumption for $\Pi(\eta,L)$ given in Eq.~\eqref{eq:Pi}.
We have done this test for all systems investigated. Results
for bulk and slit-pore systems at $T^{*}=0.5$ are
shown in Fig.~\ref{fig:Fig4}. 
For both cases the data obtained for different system sizes indeed
falls on one master curve, consistent with the theory. Numerical values
for the exponent $\nu$ are given in the subsequent Sec.~\ref{sub:Further-critical-exponents}.

\subsection{Further critical exponents\label{sub:Further-critical-exponents}}

In addition to the exponent $\nu$ mentioned in Sec.~\ref{sub:percolation_threshold},
we have determined three more critical exponents. These
are the two Fisher exponents (FE) $\tau_{\text{FE}}\text{ and }\sigma_{\text{FE}}$,
and the fractal dimensionality $D$, which are related to the cluster
size distribution, $n(s;\eta)$, and the distribution of radii of
gyration, $R(s;\eta)$.\cite{Sator2003,AharonKapitulnik1983} Thus,
the exponents $\tau_{\text{FE}}\text{, }\sigma_{\text{FE}}\text{, and }D$
have geometric (rather than thermodynamic) character. Nevertheless,
one can show \cite{Sator2003} that these exponents are linked to thermodynamic quantities
such as the pressure via scaling relations.
This link can be made in the framework of the fisher droplet model \cite{Fisher1967,Fisher1967a}
of an ideal cluster gas. Here we focus on relationships between the
``geometric'' exponents. The cluster size distribution
is defined as the average number of \textit{finite} clusters of size
$s$ at packing fraction $\eta$. Close to the percolation threshold,
this quantity displays power-law behavior, i.~e.,
\begin{equation}
	n(s;\eta_{c})\sim s^{-\tau_{\text{FE}}}
	\label{eq:pow_law_ns}
\end{equation}
as $s\rightarrow\infty$. Equation~\eqref{eq:pow_law_ns} defines the
Fisher exponent $\tau_{\text{FE}}$. The second Fisher exponent, $\sigma_{\text{FE}}$,
appears when we consider the $k$-th moment $m_{k}$ of $n(s;\eta)$ 
\begin{equation}
	m_{k}(\eta)=\sum_{s=0}^{\infty}s^{n}n(s;\eta)\text{.}
	\label{eq:def_kth_moment_of_ns}
\end{equation}
Using Eq.~\eqref{eq:pow_law_ns}, we find that $m_{k}$ fulfills the
power law 
\begin{equation}
	m_{k}(\eta)\propto|\eta-\eta_{c}|^{-(k-\tau+1)/\sigma_{\text{FE}}}\text{.}
	\label{eq:pow_law_kth_moment}
\end{equation}
We now consider the distribution of radii of gyration $R(s;\eta)$.
For a fixed size $s$, the radius of gyration of a single cluster
is defined as the averaged mean squared distance of the particles
relative to the cluster's center of mass (CM) position $\boldsymbol{r}_{\text{CM}}=\sum_{i=1}^{s}\boldsymbol{r}_{i}/s$.
This implies \cite{HeyMel1989} 
\begin{equation}
	R(s;\eta)=\sqrt{\left\langle \frac{1}{s}\sum_{i=1}^{s}(\boldsymbol{r_{i}}-\boldsymbol{r}_{\text{CM}})^{2}\right\rangle }\text{.}
	\label{eq:def_radius_of_gyration}
\end{equation}
Close to the percolation threshold, the critical behavior of this
distribution is given through the power law 
\begin{equation}
	R(s;\eta_{c})\sim s^{1/D}
	\label{eq:pow_law_rs}
\end{equation}
as $s\rightarrow\infty$. We introduced the fractal dimensionality $D$
as the critical exponent of the gyration radius.
The distribution $R(s;\eta)$ also allows us to define the correlation
length characterizing the system,
\begin{equation}
	\xi=\sqrt{2\frac{\sum_{s=0}^{\infty}R(s;\eta)s^{2}n(s;\eta)}{\sum_{s=0}^{\infty}s^{2}n(s;\eta)}}\text{.}
	\label{eq:def_cor_length}
\end{equation}
The correlation length diverges at the percolation threshold as \cite{AharonKapitulnik1983,IntPerThe}
\begin{equation}
	\xi\propto|\eta-\eta_{c}|^{-\nu},
	\label{eq:pow_law_cor_length}
\end{equation}
which defines the exponent $\nu$ already mentioned in Sec.~\ref{sub:percolation_threshold}.
Finally, using the definition \eqref{eq:def_cor_length} and the power-law
relations \eqref{eq:pow_law_kth_moment},~\eqref{eq:pow_law_rs}~and~\eqref{eq:pow_law_cor_length}
one can find the scaling relation 
\begin{equation}
	D=\frac{1}{\sigma_{\text{FE}}\nu}\text{,}
	\label{eq:connection_between_dimensionality_and_sigma_and_tau}
\end{equation}
which we use to determine the fisher exponent $\sigma_{\text{FE}}$
from the knowledge of $\nu$ and $D$. The power laws given in Eqs.~\ref{eq:pow_law_ns}~and~\ref{eq:pow_law_rs}
hold only in the limit $s\rightarrow\infty$. In fact, for the cluster size distribution the exponent $\tau_{\text{FE}}$
determines only the leading term in a series expansion of
the form
\begin{equation}
	n(s;\eta_{c})\sim s^{-\tau_{\text{FE}}}(1+A\, s^{-\Omega}+\ldots)\text{.}
	\label{eq:def_corr_to_scaling_exp}
\end{equation}
In Eq.~\eqref{eq:def_corr_to_scaling_exp} the critical exponent $\Omega$
is introduced as a first order correction to the scaling ansatz.\cite{Ziff2011}

\begin{figure}[h]
	\includegraphics[]{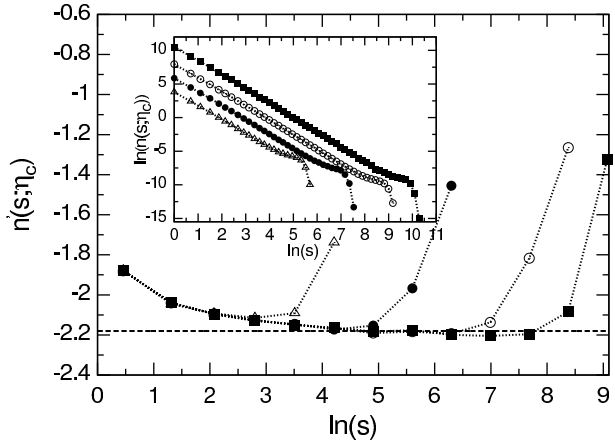}
	\caption{Local slope $n'(s;\eta_{c})$ of the double-logarithmic representation
		of the cluster-size distribution $n(s;\eta_{c})$ (inset) for the
		bulk system at reduced temperature $T^{*}=0.5$ and four system sizes
		$L=9.35\sigma\,(\vartriangle)\text{, }18.71\sigma\,(\bullet)\text{, }37.42\sigma\,(\circ)\text{ and }55.13\sigma\,(\blacksquare)$.
		The logarithmic-binning (See Sec.~\ref{sub:Further-critical-exponents})
		is done using $b=2$ for the local slope and $b=1.3$ for the inset.
		The dashed line denotes the value of the fisher exponent $\tau_\text{FE}$ for
		3D random percolation. \label{fig:Fig5}}
\end{figure}

To treat the system of finite size, we performed simulations for each analyzed
slit-pore system, using the appropriate values of the critical activity
to obtain $\eta_{c}$ in each case. We prepared the raw data by ``logarithmic
binning'' of the distributions $n(s;\eta_{c})$ and $R(s;\eta_{c})$.
To this end, we average the distributions over intervals $I_{n}=[b_{0}\frac{b^{n+1}-1}{b-1},b_{0}\frac{b^{n+2}-1}{b-1}]$,
where the width $b_{n}$ of each $I_{n}$ grows as $b_{n}=b_{0}b^{n}$
and hence produces equally spaced nodes on a logarithmic axis.\cite{Milojevic2010}
This introduces some arbitrariness via the choice of $b$. For practical
reasons we have to choose $b$ such that the noise-level of the averaged
distributions is sufficiently small. On the other hand, we must be
careful not to obscure the main tendencies. We then plot
the averaged data in a double-logarithmic representation and approximate
the local slope by difference quotients. For sufficiently
large $s$ we expect to observe a clear plateau. Indeed, from the
definition of the critical exponents $\tau_{\text{FE}}$~and~$D$
in Eqs.~\eqref{eq:pow_law_rs}~and~\eqref{eq:def_corr_to_scaling_exp}
it follows that 
\begin{equation}
	\lim_{s\rightarrow\infty}n'(s;\eta_{c})=-\tau_{\text{FE}}
	\label{eq:limit_tau}
\end{equation}
and analogously
\begin{equation}
	\text{ \ensuremath{\lim_{s\rightarrow\infty}}}R'(s;\eta_{c})=\frac{1}{D}\text{.}
	\label{eq:limit_dim}
\end{equation}
Here, $f'(s)=\partial\ln f(s)/\partial\ln s$ (where~$f=n\text{ or }R$).
In Fig.~\ref{fig:Fig5}
we show exemplary data for the functions $n'(s;\eta_{c})$ and $n(s;\eta_{c})$
(inset) for the bulk system at $T^{*}=0.5$ for four system sizes.
For the largest system, $L=55.13\sigma$, we can indeed clearly identify
a plateau. The deviations for large $s$ are due to the limitations
in the system size. For clusters with a characteristic length approaching
the system length $L$, the probability of being \textit{not} percolated
has to drop rapidly, since percolated clusters do not contribute to
the distributions.

\begin{figure}[h]
	\includegraphics[]{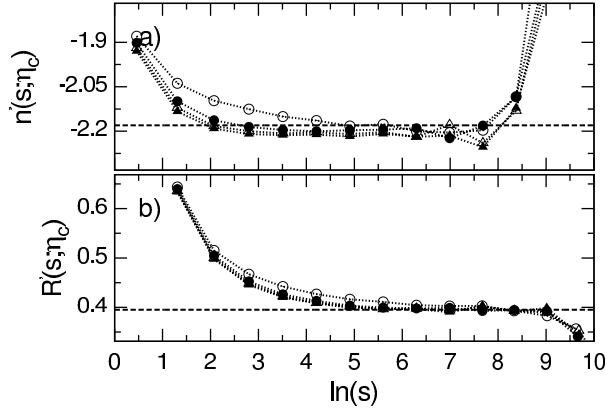}
	\caption{Logarithmic slopes $n'(s;\eta_{c})$ a) and $R'(s;\eta_{c})$ b)
		for the bulk system at the four investigated temperatures $T^{*}=0.5\,(\circ)\text{, }1.0\,(\bullet)\text{, }1.5\,(\bigtriangleup)\text{ and }2.0\,(\blacktriangle)$
		directly at the corresponding percolation thresholds. The results
		are for the largest system size investigated for the bulk system ($L=55.13\sigma$).
		The bold dashed lines denote the values for the critical
		exponent $\tau_{\text{FE}}$ (part a)) and $D$ (part b)) for 3D random
		percolation.
		\label{fig:Fig6}}
\end{figure}

In Fig.~\ref{fig:Fig6} we show, again for the bulk system,
the two functions $n'(s;\eta_{c})$ and $R'(s;\eta_{c})$ at four
different temperatures (and the largest $L$ considered).
All curves exhibit a plateau, which is reached the faster (in terms
of the cluster size $s$) the higher $T^{*}$. The important point,
however, is that the plateau heights are essentially independent of
$T^{*}$. This means that the exponents $\tau_{\text{FE}}$~and~$D$
{[}see Eqs.~ \eqref{eq:limit_tau}~and~\eqref{eq:limit_dim}{]}
extracted from our simulations may be considered as ``universal'',
that is, not influenced by the temperature. Furthermore,
the observation that the temperature does not affect the value of
the critical exponents but the speed of convergence may offer an explanation
for the observed temperature dependence of the critical exponents
for more complex interaction models.\cite{Skvor2009} An
overview of our results for the bulk system is given in Tab.~\ref{tab:critical_values_bulk}.
\begin{table}[h]
	\caption{Critical parameters for the bulk system.
		\label{tab:critical_values_bulk}}
	\begin{tabular}{ccccccc}
		\hline
		\hline 
		$T^{*}$ & $\eta_{c}$ & $\Pi_{\text{c}}$ & $\nu$ & $D$ & $\tau_{\text{FE}}$ & $\sigma_{\text{FE}}$\tabularnewline
		\hline 
		$0.5$ & $0.23165(5)$ & $0.47$ & $0.86$ & $2.48$ & $2.19$ & $0.47$\tabularnewline 
		$1.0$ & $0.32960(5)$ & $0.44$ & $0.87$ & $2.51$ & $2.20$ & $0.46$\tabularnewline
		$1.5$ & $0.36337(5)$ & $0.46$ & $0.86$ & $2.52$ & $2.21$ & $0.46$\tabularnewline 
		$2.0$ & $0.3800(3)$ & $0.48$ & $0.88$ & $2.51$ & $2.21$ & $0.45$\tabularnewline
		3D r.~p.\footnote{Literature values for the critical exponents relevant in 3D random percolation
		(r.~p.).\cite{IntPerThe,Lorenz1998}} & - & - & $0.88$ & $2.53$ & $2.18$ & $0.45$\tabularnewline
		\hline
		\hline
	\end{tabular}
\end{table}
It is visible that the critical exponents $\tau_{\text{FE}}$~and~$D$
are close to those obtained for 3D random percolation (see horizontal
lines in Fig.~\ref{fig:Fig6}).

\begin{figure}[h]
	\includegraphics[]{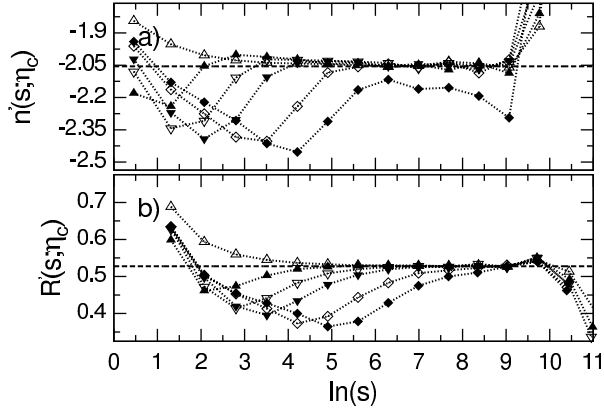}
	\caption{Same as Fig.~\ref{fig:Fig6}, but for slit-pore systems
		at temperature $T^{*}=0.5$. To improve the visibility, only the results
		for $L_{z}=1.5\sigma\,(\bigtriangleup)$, $3.0\sigma\,(\blacktriangle)$, $4.0\sigma\,(\bigtriangledown)$,
		$5.0\sigma\,(\blacktriangledown)$, $7.5\sigma\,(\lozenge)$, and $10.0\sigma\,(\blacklozenge)$
		are shown. The results are obtained for the largest system size we
		investigated in each case. The bold dashed lines denote the
		values for the critical exponent $\tau$ (part a)) and $D$ (part
		b)) for 2D random percolation.
		\label{fig:Fig7}}
\end{figure}

\begin{figure}[h]
	\includegraphics[,angle=0]{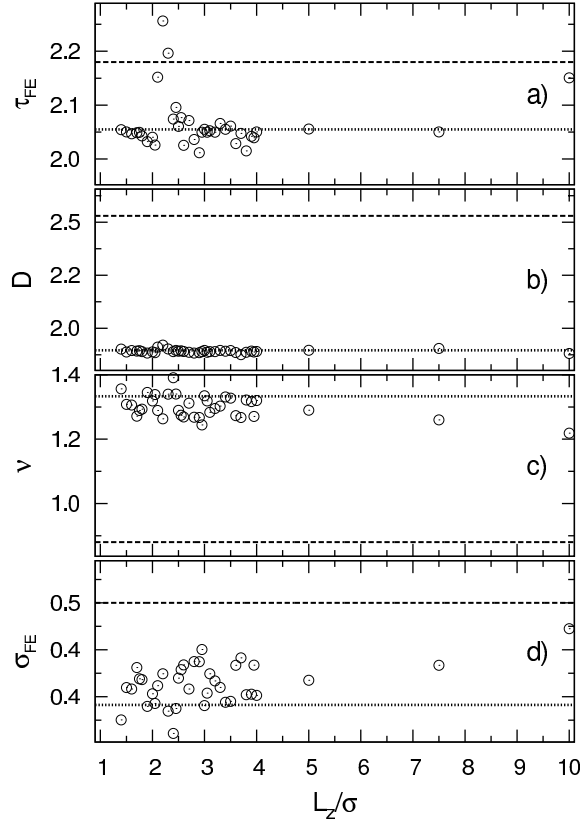}
	\caption{Estimated critical exponents $\tau_{\text{FE}}$ (a), $D$ (b), $\nu$
		(c) and $\sigma_{\text{FE}}$ (d) for the slit-pore systems at $T^{*}=0.5$
		as functions of the wall distances $L_{z}$. The exponent $\sigma_{\text{FE}}$
		is deduced from Eq.~\eqref{eq:connection_between_dimensionality_and_sigma_and_tau}.
		The lines denote the values for 3D (dashed) and 2D (dotted) random
		percolation.
		\label{fig:Fig8}}
\end{figure}

We now turn to the slit-pore systems. In Fig.~\ref{fig:Fig7}
we present $n'(s;\eta_{c})$ and $R'(s;\eta_{c})$ as functions of
$\ln\!(s)$ for six values of the wall separation $L_{z}$.
The convergence of the data (as characterized by
the appearance of a plateau) strongly depends on $L_{z}$. Specifically,
the larger $L_{z}$, the larger cluster sizes are needed before $n'$
and $R'$ reach a plateau. Moreover, in some cases, there is no clear
plateau at all (see e.g. data for $n'$ at $L_{z}=10\sigma$). For
the other cases, the observed plateau heights are close to those predicted
by 2D random percolation theory. The latter are summarized in Table~\ref{tab:critical_values_conf}.
\begin{table}[h]
	\caption{Critical exponents for 2D random percolation.\label{tab:critical_values_conf}}
	\begin{tabular}{ccccc}
		\hline
		\hline 
		 & $\nu^{\text{2D}}$ & $D^{\text{2D}}$ & $\tau_{\text{FE}}^{\text{2D}}$ & $\sigma_{\text{FE}}^{\text{2D}}$\tabularnewline
		\hline 
		exact & $\frac{4}{3}$ & $\frac{91}{48}$ & $\frac{187}{91}$ & $\frac{36}{91}$\tabularnewline
		numerical & $1.33$ & $1.90$ & $2.05$ & $0.40$\tabularnewline
		\hline
		\hline 
	\end{tabular}
\end{table}
An overview of our numerical data for the exponents $\tau_{\text{FE}}\text{, }D\text{, }\nu\text{, and }\sigma_{\text{FE}}$
as functions of the wall separation is given in Fig.~\ref{fig:Fig8}.
The graphs also indicate the corresponding literature values for
2D and 3D random percolation. Within the entire range investigated
($L_{z}<10\sigma$) the fractal dimensionality is very close to the
2D value, and a similar consistency is found for the exponent of the
correlation length, $\nu$, and the exponent $\sigma_{\text{FE}}$.
Compared to that, the data for $\tau_{\text{FE}}$ are more scattered,
consistent with what is seen in Fig.~\ref{fig:Fig7}a).

\begin{figure}[h]
	\includegraphics[]{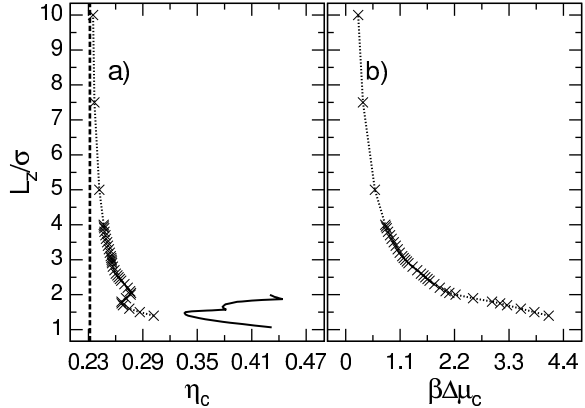}
	\caption{Relation between the percolation threshold and the wall separation
		for the slit-pore systems at $T^{*}=0.5$. Specifically, part~a)
		shows the packing fraction $\eta_{c}(L_{z})$, while b) shows the
		difference of the chemical potential to the corresponding bulk value
		$\beta\Delta\mu_{c}=\beta(\mu_{c}(L_{z})-\mu_{c}(\infty))$. The
		dotted line is a guide to the eye. The dashed line denotes the
		value of the bulk threshold at $\eta_{c}(\text{\ensuremath{\infty})}$
		{[}see~Tab.~\ref{tab:critical_values_conf}{]}. The solid
		line marks the freezing line of a pure hard sphere system which is
		confined between two hard walls (data taken from Fig.~{}4~of~Ref.~\onlinecite{Schmidt1997}).
		\label{fig:Fig9}}
\end{figure}

As far as the slit-pore systems are considered, there are (at least)
two sources of errors which come into play. The larger the wall separation
$L_{z}$, the larger cluster sizes $s$ are needed to observe a clear
plateau. Hence, our system size may still be chosen too small. Furthermore,
we can observe real 2D behavior only if we know the percolation thresholds
$\eta_{c}$ with sufficient accuracy. We expect the error tolerance
of $\eta_{c}$ to decrease with increasing wall distance $L_{z}$
\cite{Vink2006,Vink2006a} and in this spirit, we only measure effective
exponents. At $L_{z}\thickapprox2\sigma$ we observe, especially for
the fisher exponent $\tau_{\text{FE}}$, strong deviations from the
2D value of random percolation. We suppose this irregularity to occur
due to the vicinity a freezing transition, which position in the phase
diagram is known to depend on the wall separation \cite{Schmidt1997}
(see~Fig.~\ref{fig:Fig9}~a).

\subsection{Confinement-induced shift of the percolation threshold\label{sub:Confinement-induced-shift}}

So far we have focused on the critical exponents characterizing the
percolation transition. We now consider, for the confined
systems, the actual location of this transition in the phase diagram.
To this end, we depict in Fig.~\ref{fig:Fig9}~a) the critical packing
fraction $\eta_{c}$ and in Fig.~\ref{fig:Fig9}~b) the corresponding
chemical potential of the percolation threshold for various wall separations.
For comparison, we also show
the freezing line of a pure hard sphere system confined between two
parallel, hard walls (solid line) in Fig.~\ref{fig:Fig9}~a). The data was taken from
Fig.~{}4~of~Ref.~\onlinecite{Schmidt1997}. The pure hard-sphere system corresponds
to our model at large temperatures, i.~e., $T^{*}\rightarrow\infty$.

Inspecting Fig.~\ref{fig:Fig9}~a) we find that with decreasing
$L_{z}$ the percolation threshold $\eta_{c}$ increases
until $L_{z}\approx2\sigma$. At even smaller wall separations the
relation between $\eta_{c}$ and $L_{z}$ becomes non-monotonic.

In the following we focus on the behavior observed for $L_{z}\gtrsim2\sigma$,
i.~e., the shift of the percolation threshold with respect to the bulk
values. In particular, we give an argument which has successfully been 
used to explain the shifts of the gas-liquid critical point in a slit-pore
confinement.\cite{Vink2006,Vink2006a} The most remarkable point
of this argument is that it does not rely on any detailed information
of the underlying system of interest, but only on the dimensionality
and the interaction range. In Sec.~\ref{sub:percolation_threshold}
we have derived the percolation thresholds for fixed $L_{z}$ by performing
a finite-size analysis with respect to the lateral dimensions $L_{x}=L_{y}=L$
{[}see~Eq.~\eqref{eq:def_eta_av}{]}, i.~e., we have let
$L\rightarrow\infty$. This way we obtaine the percolation
threshold $\eta_{c}$, which still depends on the remaining length
$L_{z}$. We now see that, for large $L_{z}$,
the behavior upon further increasing $L_{z}$ towards infinity is
the same as that seen for $L\rightarrow\infty$ for the
bulk system.\cite{Cardy1996} In other words, we assume
that a relation similar to Eq.~\eqref{eq:def_eta_av} should also
hold for $\eta_{c}(L_{z})$. Moreover, we assume that, for sufficiently
large $L_{z}$, the corresponding exponent $\nu$ should be that of
the 3D system, $\nu^{\text{3D}}\approx0.88$. These considerations
lead to
\begin{equation}
	\Delta\eta_{c}^{*}(L_{z})=\eta_{c}^{*}(L_{z})-\eta_{c}^{*}(\infty)\propto L_{z}^{-1/\nu_{\text{3D}}}\text{.}\label{eq:shift_eta}
\end{equation}
In Eq.~\eqref{eq:shift_eta}, $\eta_{c}^{*}(L_{z})$ denotes the packing
fraction of a bulk system which is in chemical equilibrium with a slit-pore
system of width $L_{z}$ (and at average packing fraction $\eta_{c}$).
Indeed, using this notation, $\eta_{c}^{*}(\infty)$ is identical
to the percolation threshold of the bulk system $\eta_{c}^{\text{3D}}$
at $T^{*}=0.5$.

\begin{figure}[h]
	\includegraphics[]{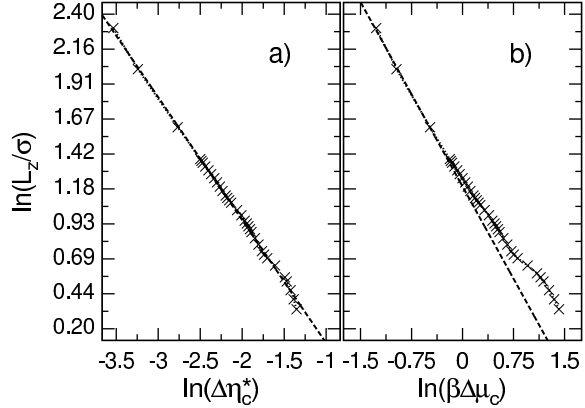}
	\caption{Scaling of the confinement-induced shifts of the percolation threshold
		at $T^{*}=0.5$. The shifts are given in terms of a) the corresponding
		densities of a bulk system in equilibrium with the slit-pore $\Delta\eta_{c}^{*}$,
		and b) the change of the chemical potential $\beta\Delta\mu_{c}$.
		The gradient of the straight lines is set to $-1/\nu_{\text{3D}}$
		as required by Eqs.~\eqref{eq:shift_eta}~and~\eqref{eq:shift_mu}.
		\label{fig:Fig10}}
\end{figure}

Since we consider our system to be above the gas-liquid critical temperature
the density is a continuous function of the chemical potential. Hence
we may write, for small deviations $\Delta\mu_{c}(L_{z})$ from the
bulk value $\mu_{c}(\infty)$ (i.~e.,~sufficiently~large~$L_{z}$),
\begin{equation}
	\Delta\mu_{c}(L_{z})=\mu_{c}(L_{z})-\mu_{c}(\infty)\propto L_{z}^{-1/\nu_{\text{3D}}}\text{.}
	\label{eq:shift_mu}
\end{equation}
In Fig.~\ref{fig:Fig10}
we compare our numerical results for $\Delta\eta_{c}^{*}(L_{z})$
and $\Delta\mu_{c}(L_{z})$ with the predictions of Eqs.~\eqref{eq:shift_eta}~and~\eqref{eq:shift_mu},
by plotting a straight line (dashed) with a fixed slop $-1/\nu_{\text{3D}}$
together with the data. From Fig.~\ref{fig:Fig10}~a) we find
that Eq.~\eqref{eq:shift_eta} is at least qualitatively fulfilled
even for the smallest wall distances investigated. Moreover, the shift
of the chemical potential shown in Fig.~\ref{fig:Fig10}~b)
is sufficiently well described by Eq.~\eqref{eq:shift_mu} for $L_{z}\gtrsim4\sigma$.
For not too small $L_{z}$ we can thus conclude that the confinement-induced
shift of the percolation threshold $\eta_{c}$ can successfully be 
described by Eqs.~\eqref{eq:shift_eta}~and~\eqref{eq:shift_mu}.
Hence we have shown that the same scaling arguments, which are usually
applied to the shift of the gas-liquid critical point in a slit-pore
confinement,\cite{Vink2006,Vink2006a} also apply to the shift of
the percolation threshold.

\section{Conclusion\label{sec:Conclusion}}

Based on grand-canonical MC simulations and a finite-size-scaling
analysis we have studied the percolation behavior of a model colloidal
suspension with an ultra-short range of attraction in three dimensions
and in slit-pore geometries. Our aim in this study was twofold: On
the one hand, we wanted to provide ``fresh'' data on the somewhat
controversial issue of the universality of critical exponents of percolation
in 3D continuous systems. On the other hand, we aimed to explore the
impact of spatial confinement on the percolation transition; a question
which was so far essentially unexplored. The model we have studied
consists of hard spheres with a square-well attraction extending over
(only) four percent of a particle diameter. As a result, the bulk
systems lacks a stable liquid phase (i.~e., the vapor-liquid-transition
predicted by a mapping procedure from the Baxter model is metastable),
and the only true transition within the fluid phase is the percolation
transition. Using a finite size analysis we have first determined
the percolation threshold, $\eta_{c}$, of the bulk system at four
temperatures above the critical temperature. Consistent with an earlier
simulation study of a similar square-well fluid (yet with a significantly
larger attraction range),\cite{J.Largo2008} we found that $\eta_{c}$
increases with increasing reduced temperature and slowly converges
to the limiting value determined by a pure hard sphere system (i.
e., the limit $T^{*}\rightarrow\infty$ of our model).

Regarding the critical exponents of the 3D percolation transition,
however, our present results differ from the earlier ones in Ref.~\onlinecite{Skvor2007}.
In that study, the exponent $\nu$ of the correlation length was found
to depend on the temperature, specifically $\nu$ decreased with $T^{*}$.
This led the authors of Ref.~\onlinecite{Skvor2007} to conclude that the exponent
is ``non-universal''. Contrary to these results, our present data
reveal no systematic dependence of $\nu$ on the temperature; instead,
the numerical values of $\nu$ fluctuate around the corresponding
value predicted by 3D random percolation theory. The question remains 
whether this contradiction could be due to the different
range of attraction considered in our study and in Ref.~\onlinecite{Skvor2007}.

In addition to $\nu$, we also studied the exponents $\tau_{\text{FE}}$
and $D$ governing the cluster size distribution $n(s;\eta_{c})$
and the radii of gyration, respectively. For sufficiently large system
sizes, we find these exponents to be independent on the temperature
and their magnitude to be very close to those predicted by 3D random percolation
theory again.

To get an insight of the impact of spatial confinement, we performed
GCMC simulations at several wall separations in the range $1.5\sigma\leq L_{z}\leq10\sigma$
at fixed reduced temperature $T^{*}=0.5$. All of these confined systems
displayed a clear percolation transition, with the percolation value
of the chemical potential (and also the resulting $\eta_{c}$) being
shifted relative to its bulk value (at $T^{*}=0.5$).

Specifically, upon decreasing $L_{z}$ from infinity (bulk limit),
we find a \textit{monotonic} increase of $\eta_{c}$ until the wall
separation comes close to the 2D limit ($L_{z}/\sigma\gtrsim2$).
Moreover, in this range of wall separations the shift of $\eta_{c}$
can be mathematically described by simple scaling arguments similar
to those applied in the framework of cross-over scaling for vapor-liquid
critical points in slit pores.\cite{Vink2006,Vink2006a} In addition
to the percolation threshold, we have also analyzed the critical exponents
for the slit-pore systems (see~Fig.~\ref{fig:Fig8}).
The exponent $\nu$ is always close to the value predicted by 2D
random percolation theory and shows no systematic dependence on $L_{z}$.
Regarding the other exponents, we have found that convergent results
(extractable from a plateau in the derivatives of the relevant distribution
functions) are generally the harder to obtain the larger the wall
separation is. Nevertheless, in the cases where we could extract reliable
data, the exponents are comparable to those of 2D random percolation.

We thus conclude that the percolation transition of colloidal systems
with ultra-short ranged interactions falls into the universality class
of random percolation, the only important factor being \textit{that} the spatial
dimension along which the system is finite (3D or 2D).

From an experimental and applicational point of view one is often
interested in the consequences of percolation, such as a strong increase
of the macroscopic conductivity, rather than in the transition itself.
In this context, we expect our results for the confined systems to be particularly
relevant, since real experimental situations often involve confinement.
We note that the model considered here is appropriate for a broad
range of colloidal suspensions involving depletion ``agents'' which
cause the short-range attraction.
From a more conceptual point of view, it is interesting that colloidal
systems with such ``sticky'' interactions --- at least in 3D --- can
not only form percolated static structures, but an also display non-trivial
dynamical behavior such as gelation and glass formation.\cite{Sciortino2011,Saika-Voivod2011,Rovigatti2012}
In particular, for spatially confined systems the interplay between
gelation and percolation is essentially unexplored. These issues will
be subject of future studies.

\end{document}